# iPTF ARCHIVAL SEARCH FOR FAST OPTICAL TRANSIENTS

Anna Y. Q. Ho,[1] S. R. Kulkarni,[1] Peter E. Nugent,[2] Weijie Zhao,[3] Florin Rusu,[3] S. Bradley Cenko,[4, 5] Vikram Ravi,[1] Mansi M. Kasliwal,[1] Daniel A. Perley,[6] Scott M. Adams,[1] Eric C. Bellm,[7] Patrick Brady,[8] Christoffer Fremling,[1] Avishay Gal-Yam,[9] David Alexander Kann,[10] David Kaplan,[8] Russ R. Laher,[11] Frank Masci,[11] Eran O. Ofek,[9] Jesper Sollerman,[12] and Alex Urban[13]

[1]*Cahill Center for Astrophysics, California Institute of Technology, MC 249-17, 1200 E California Boulevard, Pasadena, CA, 91125, USA*

[2]*Lawrence Berkeley National Laboratory, 1 Cyclotron Road, Berkeley, CA, 94720, USA*

[3]*University of California Merced, 5200 Lake Rd, Merced, CA 95340*

[4]*Astrophysics Science Division, NASA Goddard Space Flight Center, Mail Code 661, Greenbelt, MD 20771, USA*

[5]*Joint Space-Science Institute, University of Maryland, College Park, MD 20742, USA*

[6]*Astrophysics Research Institute, Liverpool John Moores University, IC2, Liverpool Science Park, 146 Browlow Hill, Liverpool L3 5RF, UK*

[7]*University of Washington Astronomy Dept., Box 351580, Seattle, WA 98195, USA*

[8]*Department of Physics, University of Wisconsin-Milwaukee, Milwaukee, WI 53201, USA*

[9]*Benoziyo Center for Astrophysics, Weizmann Institute of Science, 76100 Rehovot, Israel*

[10]*Instituto de Astrofísica de Andalucía (IAA-CSIC), Glorieta de la Astronomía s/n, 18008 Granada, Spain*

[11]*Infrared Processing and Analysis Center, California Institute of Technology, Pasadena, CA 91125, U.S.A.*

[12]*Department of Astronomy and The Oskar Klein Centre, AlbaNova University Center, Stockholm University, SE-106 91 Stockholm, Sweden*

[13]*LIGO Laboratory, California Institute of Technology, Pasadena, CA 91125, U.S.A.*

## ABSTRACT

There has been speculation of a class of relativistic explosions with an initial Lorentz factor $\Gamma_{\mathrm{init}}$ smaller than that of classical Gamma-Ray Bursts (GRBs). These "dirty fireballs" would lack prompt GRB emission but could be pursued via their optical afterglow, appearing as transients that fade overnight. Here we report a search for such transients (transients that fade by 5-$\sigma$ in magnitude overnight) in four years of archival photometric data from the intermediate Palomar Transient Factory (iPTF).

ah@astro.caltech.edu

Our search criteria yielded 50 candidates. Of these, two were afterglows to GRBs that had been found in dedicated follow-up observations to triggers from the *Fermi* GRB Monitor (GBM). Another (iPTF14yb; Cenko et al. 2015) was a GRB afterglow discovered serendipitously. Eight were spurious artifacts of reference image subtraction and one was an asteroid. The remaining 38 candidates have red stellar counterparts in external catalogs. The photometric and spectroscopic properties of the counterparts identify these transients as strong flares from M dwarfs of spectral type M3-M7 at distances of $d \approx 0.15$–$2.1$ kpc; three counterparts were already spectroscopically classified as late-type M stars. With iPTF14yb as the only confirmed relativistic outflow discovered independently of a high-energy trigger, we constrain the all-sky rate of transients that peak at $m = 18$ and fade by $\Delta m = 2\,\mathrm{mag}$ in $\Delta t = 3\,\mathrm{hr}$ to be $680\,\mathrm{yr}^{-1}$ with a 68% confidence interval of 119–$2236\,\mathrm{yr}^{-1}$. This implies that the rate of visible dirty fireballs is at most comparable to that of the known population of long-duration GRBs.

## 1. INTRODUCTION

The focus of this letter is fast (significant fading in $\lesssim 1\,\mathrm{night}$) optical transients. The sky is poorly characterized on these timescales, in part because a short cadence comes at the cost of a decrease in sky coverage. These difficulties are exacerbated by the need for rapid follow-up. By contrast, novae and supernovae (SNe) evolve on timescales of days to weeks. It is therefore not surprising that they are the best-characterized classes of transients in the optical sky.

The dominant population of fast optical transients (FOTs) is flares from Galactic low-mass main sequence stars, particularly M dwarfs (e.g., Kulkarni & Rau 2006; Rau et al. 2008; Berger et al. 2013). These flares are thought to arise from magnetic reconnection events in convective envelopes. Behind this foreground of stellar flares is a population of extragalactic relativistic explosions: the optical afterglows to gamma-ray bursts (GRBs).[1] GRBs can be explained by the "collapsar" model: a star of mass $M > 30\,M_\odot$ collapses to form a black hole, and the resulting accretion disk powers a jet (Piran 2004). The burst of $\gamma$ rays arises from within the jet, while the optical afterglow is synchrotron emission from the jet shocking the circumstellar medium.

Searching for optical or radio afterglows could yield relativistic explosions that are related to GRBs but lack high-energy emission. One example is the hypothesized class of "dirty fireballs" (Dermer et al. 2000): explosions with a lower $\Gamma_{\mathrm{init}}$ than those of classical GRBs but with similar $E_{\mathrm{iso}}$ (energy released per unit solid angle). Classical GRBs are "clean" in the sense that they have a very low baryon loading fraction, which enables matter to be accelerated to hyper-relativistic (initial Lorentz factor, $\Gamma_{\mathrm{init}} \gtrsim 100$) speeds. The primary motivation to consider dirty fireballs is the absence of a compelling reason for all relativistic explosions to have the requisite low baryon loading. The prompt emission from a dirty fireball would peak at energies below the range of $\gamma$-ray detectors. However, like a classical GRB, a dirty fireball would produce a rapidly fading (on-axis) optical afterglow and long-lived radio emission (Rhoads 2003).

Another class of optical afterglows that would lack prompt high-energy emission are off-axis ("orphan") afterglows (Rhoads 1997). Unlike for classical (on-axis, $\theta_{\mathrm{obs}} \lesssim 1/\Gamma_{\mathrm{init}}$) GRBs, an observer to an off-axis burst would not see the prompt high-energy emission, nor the initial afterglow. However, as the jet slows down it also expands sideways and as a result the afterglow becomes visible to an off-axis observer. While classical GRBs can be seen across the Universe due to relativistic beaming and Doppler boosting, orphan afterglows would be seen to shorter distances. However, the larger opening angle means that the solid angle of visibility is significantly larger than that of on-axis bursts (Nakar et al. 2002; Ghirlanda et al. 2015).

---

[1] In this paper we focus on transients related to long duration GRBs because–due to their higher energetics and larger volumetric rates–these events dominate the observed population of relativistic explosions. However, short GRB afterglows also produce fast optical transients that could conceivably pass our selection criteria.

Wide-field optical surveys have already demonstrated the technical capability to find optical afterglows independently of a GRB trigger. For example, iPTF14yb (Cenko et al. 2015) and ATLAS17aeu (Stalder et al. 2017; Bhalerao et al. 2017) were optical afterglows to GRBs identified via fading broadband afterglow emission; in both cases, only later was the "parent" GRB identified (*ibid*). Then there is the curious PTF11agg (Cenko et al. 2013), which had no identified high-energy counterpart but had other characteristic features of a GRB afterglow: a rapidly-fading optical source, a long-lived scintillating radio counterpart, and coincidence with a dwarf galaxy with an estimated redshift of $0.5 \lesssim z \lesssim 3.0$.

In this *Letter*, we report a search for fast optical transients in the *intermediate* Palomar Transient Factory (iPTF). This is similar to the search by Berger et al. (2013) for "fast optical transients" (defined as transients on timescales of 0.5 hr to 1 day) in 1.5 years of data from the Pan-STARRS-1 Medium Deep Survey (PS1/MDS). Relative to our search PS-1 is deeper ($10\sigma$ of 22.5 mag in the equivalent of $g$ and $r$ bands). They found 19 transients; of these, eight were most reasonably explained as main-belt asteroids at their turning points, and the remaining eleven were identified with quiescent M-dwarf counterparts. This work emphasized the importance of avoiding low ecliptic latitudes for future searches and highlighted the significant foreground of M-dwarf flares.

By focusing on fast transients, our search is sensitive to on-axis sources (dirty fireballs) and not off-axis events (orphan afterglows). The latter will be investigated in subsequent work, in which we search for transients that evolve rapidly on a timescale of days, like those in Drout et al. (2014) and Yang et al. (2017). §2 describes the survey, data, and search procedure, and §3 outlines the properties of the iPTF FOTs. In §4 we use the results of our search to constrain the rate of extragalactic FOTs. We conclude with a view to the upcoming Zwicky Transient Facility (ZTF; Bellm & Kulkarni 2017).

## 2. DATA AND CANDIDATE SELECTION

The intermediate Palomar Transient Factory (iPTF) ran from 1 January 2013 to 2 March 2017 as the successor to the Palomar Transient Factory (PTF; Law et al. 2009). iPTF used a camera with a $7.26 \deg^2$ field-of-view on the 48-inch Samuel Oschin Schmidt Telescope at Palomar Observatory (P48) and a real-time image subtraction pipeline (Cao et al. 2016) that was run at the National Energy Research Scientific Computing Center (NERSC) to search for transient and variable activity in the night sky. The iPTF transient surveys generally emphasized higher-cadence observations than the PTF surveys, making them well-suited for searches for fast-fading events.

The full set of candidates were saved in a database at NERSC, and the subset that passed human inspection were saved in the iPTF database at Caltech. Light curves could also be obtained using the PTF IPAC/iPTF Discovery Engine (PTFIDE) tool (Masci et al. 2017), although PTFIDE has only been run on a small subset of the iPTF database due to computational expense.

Significant improvements to the image differencing pipeline (see Section 2) were made on 1 February 2013. We therefore selected this as the start date for our search. We then performed our search in four steps, listed below. The motivation for (a) and (c) is that the afterglows discovered by optical surveys thus far manifest themselves as sources that fade overnight: iPTF14yb faded by $\sim 0.7$ mag/hr, ATLAS17aeu faded by $\sim 0.7$ mag/hr, and PTF11agg faded by $\sim 0.2$ mag/hr. With an initial magnitude of $r = 18$ mag, all three of these sources would become undetectable by iPTF (typical limiting mag $r \sim 20.5$ mag) within a night (14 hours, or 0.6 days). We chose to search for sources that have at least one pair of fading detections in order to accommodate the diversity of observed afterglow light curve shapes (e.g., Kann et al. 2010).

1. Query the NERSC database for candidates that have two detections[2] with magnitudes $m_1, m_2$ separated by $\Delta t$. This pair must satisfy the following criteria:

   (a) Fading ($m_2 > m_1$) within $\Delta t < 1$ night (0.6 day)

   (b) Real-bogus (RB[3]) score $\geq 0.3$

   (c) All detections confined to 1 night (0.6 day) [4]

   (d) All detections spatially coincident to within $1.5''$

   (e) No bad image or bad subtraction flags (`image_id` $> -1$, `sub_flag` $\neq$ 0)

2. Save all candidates from (1) to the iPTF database of named transients at Caltech. Many of these candidates were not in the iPTF database because they were not saved by human scanners (for example, because they fell below the RB threshold used during the survey).

3. Search the iPTF database (existing named transients as well as the ones added in step [2]) for candidates exhibiting afterglow behavior: significant fading, $m_2 > m_1$ at 5-$\sigma$.

4. For all candidates in (3), generate forced PSF photometry on the difference image using PTFIDE to confirm the significance of the fading.

Of the 14,961 sources with a pair of detections separated by $\Delta t < 0.6$ days, there were 1,371 sources with no detections outside this window. Of these non-repeating sources, there were 680 sources that were fading. Of these 680 sources, there were 50 that had significant (5-$\sigma$) fading.

Of the candidates, one has two detections arising from two separate asteroids[5] and eight are artifacts of image subtraction identified in visual inspection. Note that the rate of false positives is what one would expect from the raw classifier performance (Bloom et al. 2012). Removing the asteroids and artifacts, we have 41 candidates. In Figure 1 we show the $\Delta t = t_{end} - t_{start}$ and $\Delta m = m_{end} - m_{start}$ for these 41 candidates. For reference, we show PTF11agg as well as a sample of GRB afterglows from the literature (Kann et al. 2010) sampled between three hours and nine hours after peak. iPTF14cva and iPTF14cyb were afterglows discovered by PTF in searches of the *Fermi* GBM error regions (Singer et al. 2015); they correspond to the events GRB 140620A (Kasliwal et al. 2014) and GRB 140623A (von Kienlin 2014) respectively. Note that there were six more afterglows detected by iPTF in following up *Fermi* GBM triggers (Singer et al. 2015) but these did not pass the search criteria because they were detected late after the trigger time and thus were not fading significantly (all below 5-$\sigma$).

---

[2] If there are $> 2$ detections, there must exist a pair of detections satisfying (a) and (b)
[3] Brink et al. (2013)
[4] This eliminates periodic or repeating sources like AGN and variable stars.
[5] https://www.minorplanetcenter.net/cgi-bin/checkmp.cgi

The remaining 38 have red stellar hosts in external catalogs and can thus be identified as M-dwarf flares; we spectroscopically confirm these and discuss their properties in Section 3. Fortunately, all of the M dwarfs in our sample have red counterparts in external catalogs (described in Section 3) whereas none of the afterglows have detectable hosts. Indeed, of the 16 M-dwarf flares that were saved to the iPTF database during the survey (that is, prior to our search) 12 had red stellar counterparts in SDSS. The transients were thus readily classified as M-dwarf flares, although one was assigned for spectroscopic follow-up due to being faint ($r = 23.4$ mag).

## 3. PROPERTIES OF THE iPTF M-DWARF FLARES

Figure 2 shows the light curves of all 38 M-dwarf flares, superimposed with the two afterglows discovered in survey mode (as opposed to in follow-up to GRBs). The positions and classifications can be found in Appendix A and the spectra can be found in Appendix B.

For each candidate, a counterpart was present in the Panoramic Survey Telescope and Rapid Response System (Pan-STARRS; Chambers et al. 2016). For most (31 of 38) candidates, a counterpart was detected in WISE (Wright et al. 2010). Pan-STARRS host IDs and peak flare magnitudes are listed in Table 1, and a color-magnitude diagram based on Pan-STARRS $i$ and WISE W1 magnitudes is shown in Figure 3.

Of the 38 M dwarfs, three had spectra from the Sloan Digital Sky Survey (SDSS Collaboration et al. 2016). For eight of the sources that were accessible in the night sky while this work was conducted, we obtained host spectra using the Double Spectrograph on the 200-inch Hale telescope at Palomar and the Low Resolution Imaging Spectrometer (LRIS) on Keck.

In Table 1 we present derived properties of the flare stars. Note that this is not a complete sample of flaring M dwarfs in iPTF, because many were filtered out by the criterion of no prior of subsequent activity (criterion [c] in Section 2). To determine spectral type, we fit the spectra using the PyHammer software package (Kesseli et al. 2017). When a spectrum was not available, we used the quiescent Pan-STARRS colors and the relations in West et al. (2011) and Berger et al. (2013).

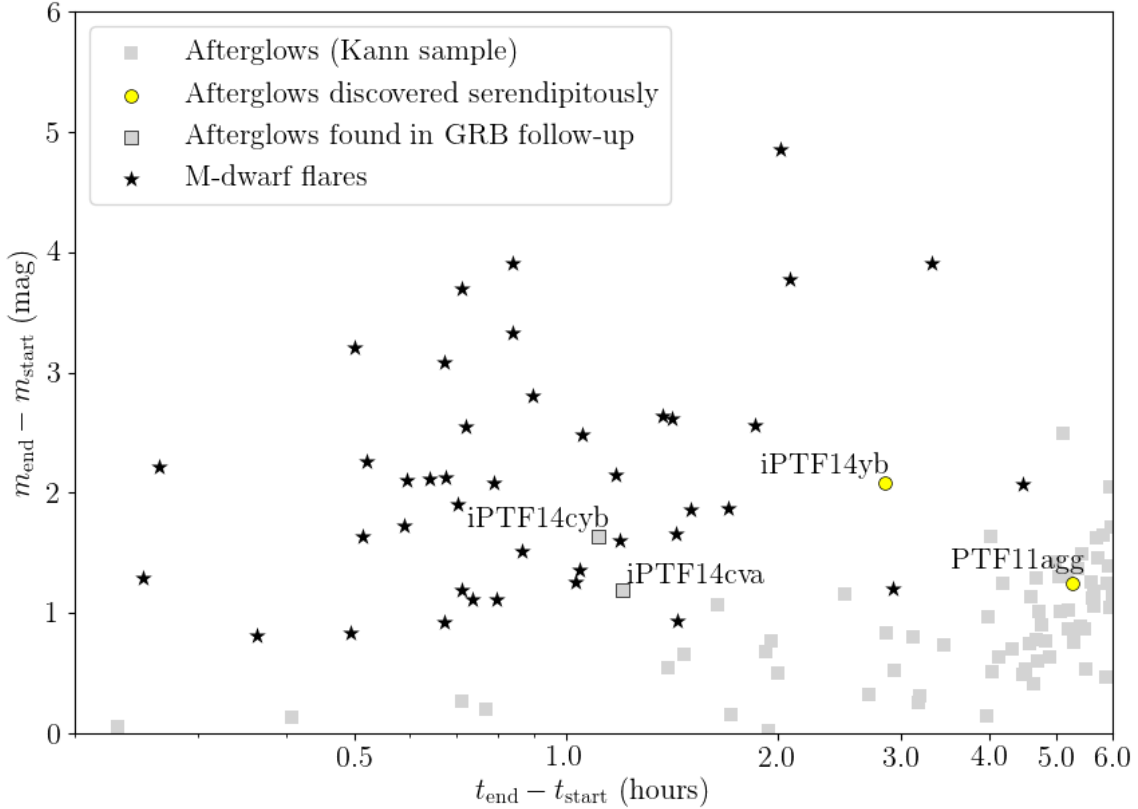

**Figure 1.** The $(\Delta t, \Delta m)$ for the 41 candidates that show significant (5-$\sigma$) intra-night fading (asteroids and artifacts of bad subtractions have been removed). The grey unlabeled points are a sample of GRB afterglows from Kann et al. (2010). For candidates with > 2 points in their light curves, we show the change in magnitude from the first observation after 3 hours to the last observation before 9 hours (times measured since the burst): $\Delta t = t_{end} - t_{start}$ and $\Delta m = m_{end} - m_{start}$. M-dwarf flares typically fade faster and are detected in Pan-STARRS (see Table 1) with a characteristic red color. Thus, in our sample, filtering out sources with red hosts exclusively identifies iPTF14yb, the GRB discovered serendipitously by iPTF, as well as two afterglows found in follow-up to Fermi GRB triggers. PTF11agg is shown for reference. There is one M-dwarf with a $\Delta t$ below the lower limit on the plot.

To estimate absolute magnitude, we used the relation between SDSS $r-z$ and $M_r$ in Bochanski et al. (2011). More precisely, we interpolated between the values in Table 5 of that paper, assuming that the stars are active and have subsolar metallicity. Because some sources are outside the SDSS footprint, we used the $r-z$ color from Pan-STARRS instead. Using the sample in our work and in Berger et al. (2013), we find that in this magnitude range ($r = 16$-$22$ mag) the $r-z$ colors are equal to within 0.1 mag between SDSS and Pan-STARRS.

Note that the values in Table 1 are subject to large uncertainties. In general, M-dwarf classifications are only reliable to within one spectral type. Taking into account uncertainties in color, metallicity, and in the interpolation tables in Bochanski et

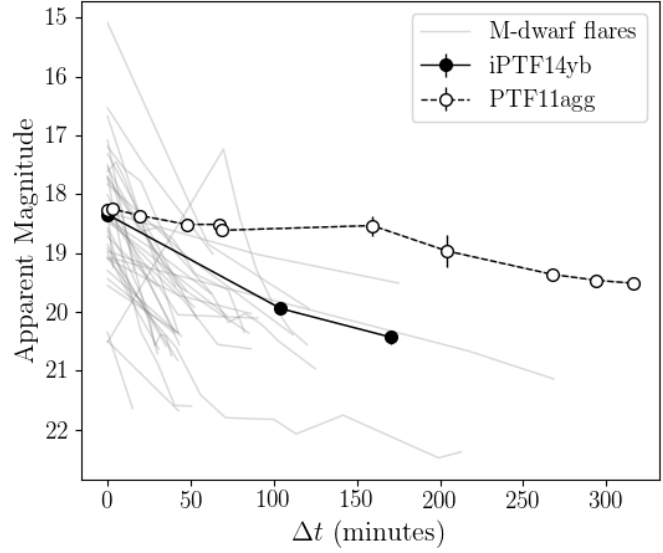

**Figure 2.** *r*-band light curves for the 38 M-dwarf flares in our sample (grey, background) overlaid with light curves of iPTF14yb and PTF11agg

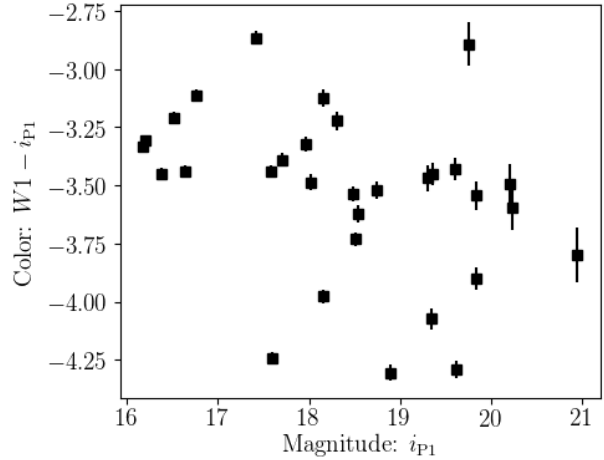

**Figure 3.** A color-magnitude diagram for 31 of the 38 M dwarfs in our sample using PanSTARRS $i$ for the magnitude and WISE $W1 -$ PanSTARRS $i$ for the color. All of the M-dwarfs have red counterparts in PanSTARRS, and most (31) have detected counterparts in WISE.

(2011), the uncertainty on absolute magnitude $M_r$ is roughly 25%. This translates into a factor of 3-4 uncertainty in distance $d$, a factor of 3-4 uncertainty in absolute height above the Galactic plane $|z|$, and an order of magnitude uncertainty in the peak luminosity of the flare $L_{\mathrm{peak,flare}}$. The $u$-band magnitude enhancement $\Delta u$ is robust to uncertainties in spectral type to within 10% and the percentile values are robust to uncertainties in spectral type to within 1%.

**Table 1**. Host and flare properties of 38 iPTF FOTs classified as M-dwarf flares. iPTF mags with a † are in $g$-band, otherwise $r$-band. In the Notes section, K means a spectrum was obtained with LRIS on Keck, P means that a spectrum was obtained with the Double Spectrograph on the Palomar 200-inch telescope, S means that an SDSS spectrum was already available. Positions and spectra can be found in the supplementary material. As described in the text, M-dwarf classifications are only reliable to within one spectral type. Other uncertainties are roughly 25% in absolute magnitude $M_r$, a factor of 3-4 in distance $d$ and absolute height above the Galactic plane $|z|$, an order of magnitude in the peak luminosity of the flare $L_{\rm peak,flare}$, 10% in the $u$-magnitude enhancement $\Delta u$, and 1% in the percentile of $\Delta u$.

| iPTF ID | PS1 ID (PSO) | $m_{\rm flare,iPTF}$ | Sp. Type | $M$ | $d$ | $|z|$ | $L_{\rm peak,flare}$ | $\Delta u$ | Percentile | Notes |
|---|---|---|---|---|---|---|---|---|---|---|
| 13agt | J170326.056+233048.207 | $20.8 \pm 0.2$ | M6 | 14.0 | 790 | 8.0 | 5.6e+30 | 4.4 | 0.99 | |
| 13asy | J122714.515+170827.218 | $20.4 \pm 0.1$ | M5 | 12.0 | 810 | 19.0 | 8.8e+30 | 3.6 | 0.97 | |
| 13bde | J163025.023+394425.607 | $20.1 \pm 0.09$ | M4 | 11.0 | 980 | 13.0 | 1.7e+31 | 4.4 | 0.99 | |
| 13bku | J132710.975+121305.263 | $20.1 \pm 0.1$ | M5 | 10.0 | 2100 | 46.0 | 8e+31 | 4.1 | 0.98 | |
| 13dqr | J022241.723+251722.567 | $21.1 \pm 0.2$ | M5 | 13.0 | 730 | 7.4 | 3.6e+30 | 4.6 | 0.99 | |
| 13gt | J133612.438+322415.839 | $20.03 \pm 0.08$ | M5 | 11.0 | 1000 | 24.0 | 1.9e+31 | 3.4 | 0.97 | |
| 13nn | J074457.731+522431.570 | $21.7 \pm 0.2$ | M5 | 11.0 | 1400 | 13.0 | 8.7e+30 | 3.7 | 0.97 | |
| 14q | J075205.876+464103.422 | $22.5 \pm 0.2^\dagger$ | M3 | 9.7 | 1800 | 16.0 | 8.1e+30 | 6.0 | 1.0 | |
| 15bgf | J204038.050+394012.906 | $20.3 \pm 0.1$ | M4 | 11.0 | 1200 | 0.46 | 1.9e+31 | 4.6 | 0.99 | |
| 15bm | J075629.265+195502.966 | $20.6 \pm 0.1$ | M7 | 14.0 | 350 | 2.4 | 1.4e+30 | 5.8 | 1.0 | |
| 15dto | J002938.210+034148.808 | $20.4 \pm 0.2^\dagger$ | M5 | 12.0 | 560 | 10.0 | 5.7e+30 | 3.0 | 0.92 | K |
| 15ell | J034044.994+181735.258 | $19.56 \pm 0.09$ | M5 | 12.0 | 1100 | 9.5 | 3.4e+31 | 4.6 | 0.99 | K |
| 16bse | J204045.160+411809.265 | $20.7 \pm 0.2$ | M4 | 11.0 | 590 | 0.062 | 3.4e+30 | 4.6 | 0.99 | P |
| 16bxw | J002145.452+005843.242 | $19.5 \pm 0.2$ | M6 | 12.0 | 180 | 3.4 | 9.9e+29 | 3.7 | 0.97 | S,P |
| 16ccd | J025954.415+602506.863 | $19.89 \pm 0.08$ | M5 | 12.0 | 740 | 0.32 | 1.2e+31 | 3.3 | 0.94 | K |
| 17ady | J141130.672+304100.846 | $21.6 \pm 0.1$ | M7 | 14.0 | 150 | 3.3 | 9.9e+28 | 3.8 | 0.97 | |
| 17ahn | J164144.856+403623.379 | $20.3 \pm 0.1$ | M5 | 10.0 | 1900 | 24.0 | 4.2e+31 | 5.0 | 0.99 | K |
| 17alz | J022942.051+191822.355 | $20.6 \pm 0.1$ | M6 | 11.0 | 480 | 5.5 | 4e+30 | 5.3 | 0.99 | P |
| 17amj | J012608.197+353352.587 | $20.12 \pm 0.09$ | M5 | 11.0 | 950 | 7.7 | 1.9e+31 | 4.0 | 0.98 | K |
| 17bub | J054206.049+700935.192 | $19.92 \pm 0.09$ | M4 | 11.0 | 430 | 2.6 | 5.5e+30 | 3.3 | 0.94 | |
| 17eur | J080132.966+180821.586 | $19.56 \pm 0.08$ | M3 | 9.6 | 640 | 4.5 | 4.5e+30 | 4.2 | 0.98 | |
| 17hce | J020737.876+135531.430 | $20.63 \pm 0.09$ | M5 | 12.0 | 540 | 7.4 | 4.1e+30 | 5.1 | 0.99 | |
| 17hhv | J072756.444+180748.975 | $20.3 \pm 0.2$ | M4 | 11.0 | 280 | 1.4 | 3e+30 | 4.4 | 0.99 | |
| 17hmf | J093025.725+114653.074 | $19.3 \pm 0.1$ | M4 | 11.0 | 500 | 6.2 | 2e+30 | 6.3 | 1.0 | |
| 17hmz | J080557.336+154053.582 | $21.0 \pm 0.2$ | M5 | 13.0 | 240 | 1.7 | 6.8e+29 | 4.2 | 0.99 | |
| 17ipt | J133442.745+055903.060 | $20.6 \pm 0.2$ | M5 | 12.0 | 190 | 3.9 | 4.1e+29 | 4.1 | 0.98 | |
| 17iwk | J153313.078+571537.332 | $20.6 \pm 0.1$ | M4 | 10.0 | 360 | 5.4 | 1.8e+30 | 5.1 | 0.99 | |
| 17jlt | J151344.316+200736.440 | $20.4 \pm 0.1$ | M5 | 12.0 | 480 | 8.2 | 6.2e+30 | 4.6 | 0.99 | |
| 17jq | J032221.653+264423.619 | $19.63 \pm 0.09$ | M5 | 12.0 | 470 | 3.6 | 5.8e+30 | 3.8 | 0.97 | |
| 17jqb | J150608.089+134859.802 | $19.7 \pm 0.1$ | M5 | 13.0 | 620 | 11.0 | 8.2e+30 | 4.8 | 0.99 | |
| 17jvl | J114254.502+275546.738 | $19.9 \pm 0.1$ | M4 | 11.0 | 190 | 4.4 | 1.4e+30 | 4.7 | 0.99 | |
| 17knl | J083105.765+160952.079 | $19.29 \pm 0.07$ | M4 | 10.0 | 680 | 6.0 | 2.2e+31 | 5.0 | 0.99 | |
| 17mlj | J074900.627+210136.013 | $19.0 \pm 0.2$ | M4 | 11.0 | 220 | 1.5 | 1e+30 | 7.4 | 1.0 | |
| 17py | J162922.139+335645.582 | $19.9 \pm 0.2$ | M7 | 14.0 | 370 | 4.9 | 5.5e+30 | 3.5 | 0.97 | |
| 17qfn | J103422.298+091040.949 | $19.24 \pm 0.06$ | M4 | 11.0 | 610 | 9.9 | 3.8e+30 | 5.7 | 1.0 | |

**Table 1** *(continued)*

| iPTF ID | PS1 ID (PSO) | $m_{\mathrm{flare,iPTF}}$ | Sp. Type | $M$ | $d$ | $|z|$ | $L_{\mathrm{peak,flare}}$ | $\Delta u$ | Percentile | Notes |
|---------|--------------|---------------------------|----------|-----|-----|-------|---------------------------|-----------|------------|-------|
| 17rzn | J084115.859+181628.242 | $20.7 \pm 0.1$ | M6 | 14.0 | 280 | 2.8 | 3.5e+29 | 6.4 | 1.0 | S |
| 17yz  | J104639.306+323916.760 | $21.6 \pm 0.2$ | M6 | 14.0 | 200 | 3.8 | 5.6e+29 | 6.2 | 1.0 | S |
| 17ze  | J131505.985+430400.947 | $20.4 \pm 0.1$ | M5 | 12.0 | 290 | 6.6 | 1.2e+30 | 5.1 | 0.99 |   |

The fraction of active stars and the flare rate have been found to increase with later spectral type (Kowalski et al. 2009; West et al. 2011), so it is not surprising that most of the stars in our sample are spectral type M5 or M6. Furthermore, these stars are all located at small vertical distances from the Galactic plane, consistent with the finding in Kowalski et al. (2009) that flare rate decreases strongly with distance from the plane (stars lying close to the plane are younger, which may be associated with stronger activity).

Next, we compare the flare amplitudes to the sample in Kowalski et al. (2009) and list the percentile in the last column of Table 1. Kowalski et al. (2009) measure flare luminosities in $u$ band[6]. To estimate the $\Delta u$ of the flares in our sample, we convert $\Delta r$ or $\Delta g$ to $\Delta u$ using the model in Davenport et al. (2012). The flares in our sample are large compared to those from most active stars of this spectral type. This is because of our selection criteria: the typical uncertainty on an iPTF magnitude is $\sim 0.1$, so a 5-$\sigma$ change in magnitude is typically $\Delta r > 0.5$ or $\Delta g > 0.5$. A magnitude change of $\Delta r > 0.5$ corresponds to $\Delta u > 3$ in the Davenport et al. (2012) model, which is already at the 92nd percentile of the distribution in Kowalski et al. (2009).

## 4. RATE OF RELATIVISTIC FAST OPTICAL TRANSIENTS IN iPTF

With iPTF14yb remaining the only confirmed afterglow in iPTF discovered independently of a high-energy trigger, we can constrain the rate of transients that exhibit the same fading behavior (peak at $m = 18$, fade by $\Delta m = 2\,\mathrm{mag}$ in $\Delta t = 3\,\mathrm{hr}$). With our selection criteria and observations from 1 February 2013 through 2 March 2017, we follow a similar procedure to that in Section 5 of Cenko et al. (2015). We take all of the iPTF observations over this four-year period. We insert the light curve of iPTF14yb (for simplicity) stepping through a range of burst times. Using the limiting magnitude of the exposure and the brightness of the source at the time of observation, we determine whether the event would have been detected using our search criteria, i.e. two detections with a 5-$\sigma$ difference in magnitude.

This gives a total areal exposure of $A_{\mathrm{eff}} = 22,146\,\mathrm{deg^2}$ days. So, we constrain the all-sky rate of on-axis relativistic transients similar to iPTF14yb to be

$$\mathcal{R} \equiv \frac{N_{\mathrm{rel}}}{A_{\mathrm{eff}}} = \frac{1}{22,146\,\mathrm{deg^2\,day}} \times \frac{365.25\,\mathrm{day}}{\mathrm{yr}} \times \frac{41,253\,\mathrm{deg^2}}{\mathrm{sky}} = 680\,\mathrm{yr^{-1}} \qquad (1)$$

with a 68% confidence interval from Poisson statistics of $119\text{--}2236\,\mathrm{yr^{-1}}$. The expected rate of classical optical afterglows that can be detected by (i)PTF is two-thirds of the rate of on-axis *Swift* GRBs, or $\mathcal{R} = 970^{+53}_{-74}\,\mathrm{yr}^{-1}$ (Cenko et al. 2015). Thus, our search sets a limit on the relative rate of visible dirty fireballs to classical on-axis afterglows and suggests that it is at most comparable.

---

[6] M-dwarf flares are typically studied in the $u$ band because this holds the greatest contrast between the blue flare and red host.

We now estimate the volumetric rate of transients with these characteristics (peak at $m = 18$, fade by $\Delta m = 2\,\mathrm{mag}$ in $\Delta t = 3\,\mathrm{hr}$). iPTF14yb was observed at redshift $z = 1.9733 \pm 0.0003$ with spectral index $\beta = 1.3 \pm 0.1$ and apparent magnitude $m_\mathrm{p} = 18.16 \pm 0.03$ in its first discovery image. Applying a standard $K$-correction (Hogg et al. 2002), this corresponds to an absolute magnitude $M_\mathrm{p} = -27.5 \pm 0.1$ in the $r$-band some $\sim$300 s after the initial outburst, which is fairly typical of the afterglows of *Swift* long GRBs (Cenko et al. 2009). Assuming iPTF14yb represents a population of standard candles (which is not really the case; see Kann et al. 2010) an identical explosion would appear with magnitude $m \approx 21$ if it occurred at redshift $z \approx 3$; thus we infer a volumetric rate of 0.395 $\mathrm{Gpc}^{-3}\,\mathrm{yr}^{-1}$ with a $1\sigma$ credible interval of (0.022–0.708) $\mathrm{Gpc}^{-3}\,\mathrm{yr}^{-1}$ . This is roughly consistent with 1/3–2/3 the rate of long-duration *Swift* GRBs in the local universe (1.3 $\mathrm{Gpc}^{-3}\,\mathrm{yr}^{-1}$; Wanderman & Piran 2010) without accounting for beaming. A more detailed analysis of this volumetric rate is forthcoming (Urban et al., in prep.).

## 5. CONCLUSIONS

The Zwicky Transient Facility (ZTF; Bellm & Kulkarni 2017) has just achieved first light, and with its $47\,\mathrm{deg}^2$ field of view and faster readout will represent on average a 12-fold increase in volumetric survey speed over PTF. Thus, in one routine semester of ZTF, we will be able to reproduce the coverage of iPTF, setting very strong limits on the rates of extragalactic fast optical transients, or potentially providing the first confirmed detection of afterglows lacking prompt high-energy emission.

So far, it seems that M-dwarf flares are the only astrophysical contaminant in searching for afterglows via rapidly-fading emission. In particular, our selection criteria identify flares from late-type M-dwarfs in the top decile of flare amplitude. Such events are rare due to the intrinsic faintness of late-type M dwarfs and the anti-correlation of flare frequency with flare energy (e.g., Davenport 2016). Wide-area, high-cadence surveys like PTF and ZTF are thus well-suited for identifying the most extreme examples of flaring activity (so-called "hyperflares"), aiding studies of chromospheric activity and stellar dynamos.

That said, the cadence of these wide-field surveys (PTF, ZTF, LSST) is not well-suited for constraining detailed physics of flares. Instead, the cadence is more suited to flare population statistics. The spatial distribution of these extreme examples of flaring activity is interesting because flares are typically an indicator of stellar youth. The same ZTF data (and other such surveys) can be used to measure rotation rates and therefore estimate stellar ages (gyrochronology). Therefore, properly modeling the transient contribution for flares could result in a relation between activity and rotation period for these stars. The latter is usually taken as a proxy for age.

APPENDIX

A. TABLE OF iPTF FAST OPTICAL TRANSIENTS

**Table 2**. iPTF Fast Optical Transients

| PTF ID | RA | Dec | UT Date | Classification |
|--------|------|------|---------|----------------|
| 13agt | 17:03:26.07 | +23:30:48.0 | 2013-04-04 | M-dwarf |
| 13asy | 12:27:14.53 | +17:08:27.2 | 2013-05-04 | M-dwarf |
| 13bde | 16:30:25.03 | +39:44:25.5 | 2013-05-15 | M-dwarf |
| 13bku | 13:27:11.00 | +12:13:05.2 | 2013-06-01 | M-dwarf |
| 13dqr | 02:22:41.74 | +25:17:22.6 | 2013-10-04 | M-dwarf |
| 13gt | 13:36:12.43 | +32:24:15.8 | 2013-02-18 | M-dwarf |
| 13nn | 07:44:57.71 | +52:24:31.4 | 2013-03-06 | M-dwarf |
| 13qz | 12:02:07.82 | +01:22:50.8 | 2013-03-13 | Bad Subtraction |
| 14cva | 18:47:29.00 | +49:43:51.7 | 2014-06-20 | Afterglow |
| 14cyb | 15:01:53.41 | +81:11:29.0 | 2014-06-23 | Afterglow |
| 14q | 07:52:05.86 | +46:41:03.2 | 2014-01-03 | M-dwarf |
| 14ts | 10:05:47.69 | +10:25:52.2 | 2014-02-22 | Rock |
| 14yb | 14:45:58.01 | +14:59:35.3 | 2014-02-26 | Afterglow |
| 15bgf | 20:40:38.04 | +39:40:12.7 | 2015-06-12 | M-dwarf |
| 15bm | 07:56:29.27 | +19:55:02.9 | 2015-01-18 | M-dwarf |
| 15dto | 00:29:38.21 | +03:41:48.9 | 2015-11-09 | M-dwarf |
| 15ell | 03:40:45.01 | +18:17:35.2 | 2015-11-20 | M-dwarf |
| 16bse | 20:40:45.14 | +41:18:08.9 | 2016-07-11 | M-dwarf |
| 16bxw | 00:21:45.47 | -00:58:43.1 | 2013-10-01 | M-dwarf |
| 16ccd | 02:59:54.41 | +60:25:06.7 | 2016-11-23 | M-dwarf |
| 16hdn | 00:58:13.16 | +06:24:00.9 | 2016-10-13 | Bad Subtraction |
| 17ady | 14:11:30.65 | +30:41:00.7 | 2013-03-15 | M-dwarf |
| 17ahn | 16:41:44.86 | +40:36:23.0 | 2013-05-21 | M-dwarf |
| 17alz | 02:29:42.04 | +19:18:22.5 | 2013-09-04 | M-dwarf |
| 17amj | 01:26:08.20 | +35:33:52.6 | 2013-09-07 | M-dwarf |
| 17bub | 05:42:06.04 | +70:09:35.1 | 2017-03-02 | M-dwarf |
| 17eur | 08:01:32.94 | +18:08:21.3 | 2014-01-07 | M-dwarf |
| 17hce | 02:07:37.89 | +13:55:31.5 | 2014-11-17 | M-dwarf |
| 17hhv | 07:27:56.48 | +18:07:49.0 | 2015-01-15 | M-dwarf |
| 17hmf | 09:30:25.74 | +11:46:53.1 | 2015-02-21 | M-dwarf |
| 17hmz | 08:05:57.36 | +15:40:54.2 | 2015-03-10 | M-dwarf |
| 17ipt | 13:34:42.67 | +05:59:02.4 | 2013-03-14 | M-dwarf |
| 17iwk | 15:33:13.10 | +57:15:36.8 | 2013-04-22 | M-dwarf |
| 17jlt | 15:13:44.32 | +20:07:36.6 | 2013-03-15 | M-dwarf |
| 17jq | 03:22:21.67 | +26:44:23.2 | 2014-02-11 | M-dwarf |
| 17jqb | 15:06:08.11 | +13:48:59.9 | 2013-03-15 | M-dwarf |
| 17jvl | 11:42:54.48 | +27:55:46.7 | 2013-03-14 | M-dwarf |
| 17knl | 08:31:05.76 | +16:09:51.7 | 2014-05-19 | M-dwarf |
| 17mlj | 07:49:00.62 | +21:01:35.7 | 2014-01-20 | M-dwarf |

**Table 2** *(continued)*

| PTF ID | RA | Dec | UT Date | Classification |
|--------|-----|-----|---------|----------------|
| 17py | 16:29:22.15 | +33:56:45.5 | 2013-03-14 | M-dwarf |
| 17qfn | 10:34:22.32 | +09:10:40.9 | 2015-02-26 | M-dwarf |
| 17rzn | 08:41:15.87 | +18:16:28.0 | 2015-01-19 | M-dwarf |
| 17tq | 04:57:50.59 | +00:27:30.8 | 2013-12-14 | Bad Subtraction |
| 17ufp | 08:01:27.39 | +18:08:07.0 | 2015-01-19 | Bad Subtraction |
| 17uo | 07:18:12.25 | +64:21:19.6 | 2014-01-18 | Bad Subtraction |
| 17whs | 01:54:27.77 | +20:29:35.9 | 2013-10-05 | Bad Subtraction |
| 17wok | 05:15:28.12 | +01:30:47.1 | 2013-12-14 | Bad Subtraction |
| 17wsv | 08:09:42.13 | +19:45:05.3 | 2015-01-19 | Bad Subtraction |
| 17yz | 10:46:39.27 | +32:39:16.4 | 2013-03-11 | M-dwarf |
| 17ze | 13:15:06.05 | +43:04:01.5 | 2013-03-12 | M-dwarf |

## B. SPECTRA OF M-DWARF HOSTS

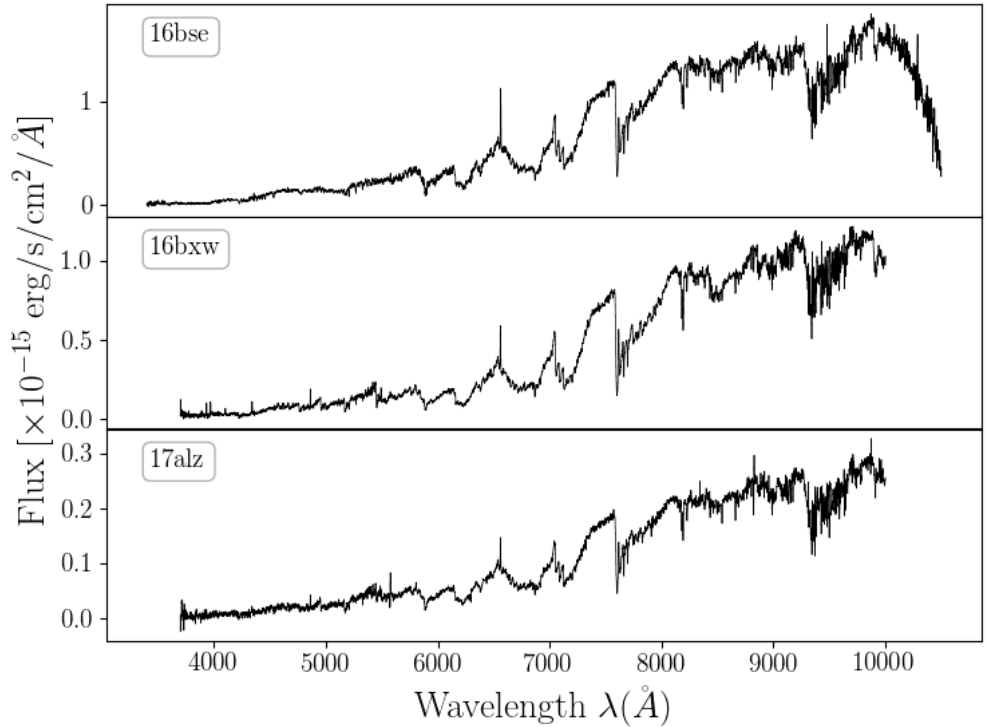

**Figure 4.** DBSP spectra of three of the M dwarfs in our sample

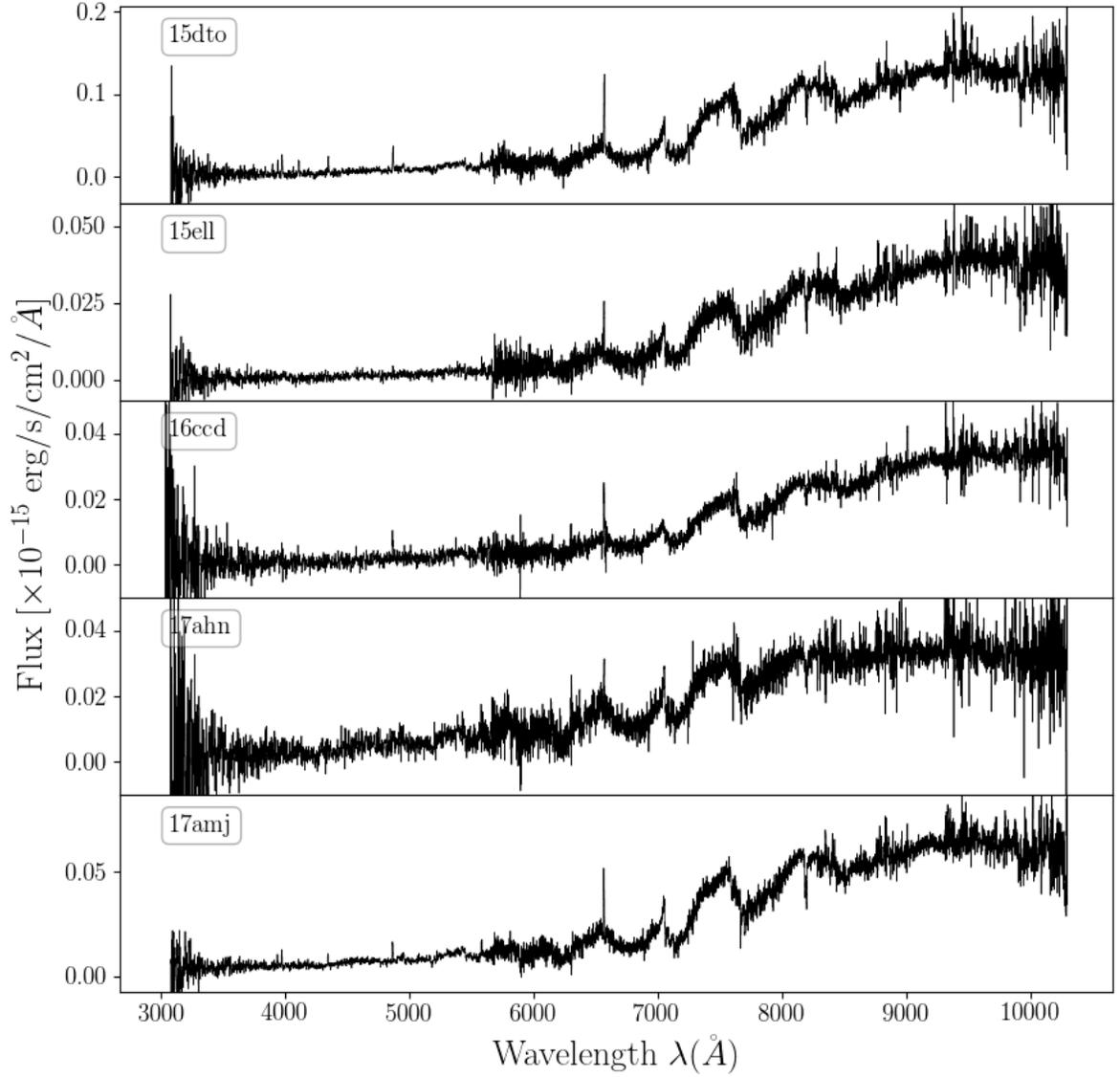

**Figure 5.** LRIS spectra of five of the M dwarfs in our sample

It is a pleasure to thank Yi Cao, Jim Davenport, Adam Miller, Yuguang Chen, Harish Vendantham, Lynne Hillenbrand, and Trevor David for helpful discussions and assistance. We are grateful to the anonymous referee for constructive feedback that improved the quality of the paper. A.Y.Q.H. was supported by a National Science Foundation Graduate Research Fellowship under Grant No. DGE1144469. DAK acknowledges support from from the Spanish research project AYA 2014-58381-P and the Juan de la Cierva Incorporacíon fellowship IJCI-2015-261. This work was supported by the GROWTH project funded by the National Science Foundation under PIRE Grant No 1545949. The Intermediate Palomar Transient Factory project is a scientific collaboration among the California Institute of Technology, Los Alamos National Laboratory, the University of Wisconsin, Milwaukee, the Oskar Klein Center, the Weizmann Institute of Science, the TANGO Program of the University System of Taiwan, and the Kavli Institute for the Physics and Mathematics of the Universe. This research made use of Astropy, a community-developed core Python package for Astronomy (Astropy Collaboration et al. 2013).

## REFERENCES

Bellm, E., & Kulkarni, S. 2017, Nature Astronomy, 1, 0071

Berger, E., Leibler, C. N., Chornock, R., et al. 2013, ApJ, 779, 18

Bhalerao, V., Kasliwal, M. M., Bhattacharya, D., et al. 2017, arXiv:1706.00024

Bloom, J. S., Richards, J. W., Nugent, P. E., et al. 2012, PASP, 124, 1175

Bochanski, J. J., Hawley, S. L., & West, A. A. 2011, AJ, 141, 98

Brink, H., Richards, J. W., Poznanski, D., et al. 2013, MNRAS, 435, 1047

Cao, Y., Nugent, P. E., & Kasliwal, M. M. 2016, PASP, 128, 114502

Cenko, S. B., Kelemen, J., Harrison, F. A., et al. 2009, ApJ, 693, 1484

Chambers, K. C., Magnier, E. A., Metcalfe, N., et al. 2016, arXiv:1612.05560

Cenko, S. B., Kulkarni, S. R., Horesh, A., et al. 2013, ApJ, 769, 130

Cenko, S. B., Urban, A. L., Perley, D. A., et al. 2015, ApJL, 803, L24

Davenport, J. R. A., Becker, A. C., Kowalski, A. F., et al. 2012, ApJ, 748, 58

Davenport, J. R. A. 2016, ApJ, 829, 23

Dermer, C. D., Chiang, J., & Mitman, K. E. 2000, ApJ, 537, 785

Drout, M. R., Chornock, R., Soderberg, A. M., et al. 2014, ApJ, 794, 23

Ghirlanda, G., Salvaterra, R., Campana, S., et al. 2015, A&A, 578, A71

Hogg, D. W., Baldry, I. K., Blanton, M. R., & Eisenstein, D. J. 2002, arXiv:astro-ph/0210394

Kann, D. A., Klose, S., Zhang, B., et al. 2010, ApJ, 720, 1513

Kasliwal, M. M., Cenko, S. B., & Singer, L. P. 2014, GRB Coordinates Network, 16425, 1

Kesseli, A. Y., West, A. A., Veyette, M., et al. 2017, ApJS, 230, 16

von Kienlin, A. 2014, GRB Coordinates Network, 16450, 1

Kowalski, A. F., Hawley, S. L., Hilton, E. J., et al. 2009, AJ, 138, 633

Kulkarni, S. R., & Rau, A. 2006, ApJL, 644, L63

Law, N. M., Kulkarni, S. R., Dekany, R. G., et al. 2009, PASP, 121, 1395

Masci, F. J., Laher, R. R., Rebbapragada, U. D., et al. 2017, PASP, 129, 014002

Nakar, E., Piran, T., & Granot, J. 2002, ApJ, 579, 699

Piran, T. 2004, Reviews of Modern Physics, 76, 1143

Rau, A., Ofek, E. O., Kulkarni, S. R., et al. 2008, ApJ, 682, 1205-1216

Rhoads, J. E. 1997, ApJL, 487, L1

Rhoads, J. E. 2003, ApJ, 591, 1097

Astropy Collaboration, Robitaille, T. P., Tollerud, E. J., et al. 2013, A&A, 558, A33

SDSS Collaboration, Albareti, F. D., Allende Prieto, C., et al. 2016, arXiv:1608.02013

Singer, L. P., Kasliwal, M. M., Cenko, S. B., et al. 2015, ApJ, 806, 52

Stalder, B., Tonry, J., Smartt, S. J., et al. 2017, arXiv:1706.00175

Wanderman, D., & Piran, T. 2010, MNRAS, 406, 1944

West, A. A., Hawley, S. L., Bochanski, J. J., et al. 2008, AJ, 135, 785

West, A. A., Morgan, D. P., Bochanski, J. J., et al. 2011, AJ, 141, 97

Wright, E. L., Eisenhardt, P. R. M., Mainzer, A. K., et al. 2010, AJ, 140, 1868-1881

Yang, S., Valenti, S., Cappellaro, E., et al. 2017, arXiv:1710.05864